# Triaxially strained suspended graphene for large-area pseudo-magnetic fields

**Manlin Luo,**[1,+] **Hao Sun,**[1,2,+] **Zhipeng Qi,**[3] **Kunze Lu,**[1] **Melvina Chen,**[1] **Dongho Kang,**[1,4] **Youngmin Kim,**[1] **Daniel Burt,**[1] **Xuechao Yu,**[1] **Chongwu Wang,**[1] **Young Duck Kim,**[5] **Hong Wang,**[1] **Qi Jie Wang,**[1,2] **and Donguk Nam,**[1, *]

[1] *School of Electrical and Electronic Engineering, Nanyang Technological University, 50 Nanyang Avenue, Singapore 639798, Singapore*
[2] *Division of Physics and Applied Physics, School of Physical and Mathematical Sciences, Nanyang Technological University, 21 Nanyang Link, Singapore 637371, Singapore*
[3] *School of Physics and Optoelectronic Engineering, Nanjing University of Information Science & Technology, Nanjing 210044, China*
[4] *School of Electrical Engineering and Computer Science, Gwangju Institute of Science and Technology, Gwangju 61005, Republic of Korea*
[5] *Department of Physics, Kyung Hee University, Seoul 02447, Republic of Korea*
[+]*these authors contributed equally to this work*
*Corresponding author: dnam@ntu.edu.sg*

**Strain-engineered graphene has garnered much attention recently owing to the possibilities of creating substantial energy gaps enabled by pseudo-magnetic fields. While theoretical works proposed the possibility of creating large-area pseudo-magnetic fields by straining monolayer graphene along three crystallographic directions, clear experimental demonstration of such promising devices remains elusive. Herein, we experimentally demonstrate a triaxially strained suspended graphene structure that has the potential to possess large-scale and quasi-uniform pseudo-magnetic fields. Our structure employs uniquely designed metal electrodes that function both as stressors and metal contacts for current injection. Raman characterization and tight-binding simulations suggest the possibility of achieving pseudo-magnetic fields over a micrometer-scale area. Current-voltage measurements confirm an efficient current injection into graphene, showing the potential of our devices for a new class of optoelectronic applications. We also theoretically propose a photonic crystal-based laser structure that obtains strongly localized optical fields overlapping with the spatial area under uniform pseudo-magnetic fields, thus presenting a practical route towards the realization of graphene lasers.**

---

The discovery of graphene has created new opportunities in various distinct research fields such as condensed matter physics and high-performance electronics and optoelectronics [1]. Despite its countless superior properties, graphene's gapless feature has been considered a major bottleneck towards creating bandgap-enabled nanoelectronic devices for switching applications [2] and nanophotonic devices possessing strong bandgap transitions [3]. Among a large variety of approaches for creating energy gaps [3–5], strain engineering has arisen as one of the strongest candidates for producing the gapped graphene [5]. For instance, it was theoretically predicted that the gapless graphene can have sizable energy gaps upon the application of more than 20% strain [5]. However, most works attempting to strain large-area monolayer graphene have reported limited strain values of less than 1.5% [6,7], thereby suppressing the hope of creating energy gaps in graphene.

It is well known that the charge carriers confined to two dimensions travel in cyclotron orbits under a strong magnetic field, resulting in the creation of energy gaps enabled by Landau quantization [8]. It was theoretically predicted in 2009 that a well-designed strain in graphene can also allow the charge carriers to behave in the same way without requiring the use of an external magnetic field [9]. This effect was termed pseudo-magnetic fields [9].

Over the past decade, the existence of pseudo-magnetic fields has been experimentally proven by scanning tunneling spectroscopy (STS) on deformed graphene sheets [10–12]. Very recently, it was demonstrated that the pseudo-magnetic fields in such deformed monolayer graphene sheets can significantly influence the hot carrier dynamics by creating large pseudo-Landau levels [13]. Unfortunately, however, the spatial area of the induced pseudo-magnetic fields in most studies reported until today is limited to the nanometer scale [10–13], which has prevented researchers from harnessing the unique pseudo-magnetic fields in optoelectronic devices because of the micrometer-scale optical diffraction limit. A few research groups have recently reported the possibility of creating large-scale pseudo-magnetic fields [14–18]. However, it remains elusive whether it is feasible to achieve spatially uniform pseudo-magnetic fields at the micrometer scale, which holds the key towards harnessing pseudo-magnetic fields for optoelectronic applications.

In this work, we present an experimental demonstration of triaxially strained suspended graphene structures that can obtain quasi-uniform pseudo-magnetic fields over a large scale. The unique design of our structure allows achieving a micrometer-scale non-uniform strain with a relatively constant strain gradient, which plays a key role in obtaining uniform pseudo-magnetic fields. Three arms of precisely patterned graphene are attached to three metal stressors, which also allow an efficient current injection into graphene. We also propose a hybrid laser structure employing a two-dimensional (2D) photonic crystal and triaxially strained graphene as an optical cavity and gain medium, respectively. Full three-dimensional (3D) finite-difference time-domain (FDTD) simulations confirm a strong optical mode overlap with an area under uniform pseudo-magnetic fields. Our results suggest a new route to realize high-performance graphene optoelectronic devices by harnessing large-scale uniform pseudo-magnetic fields in uniquely strained graphene.

Figures 1(a) and (b) show top- and tilted-view scanning electron microscopy (SEM) images of the fabricated device, where a fully suspended graphene sheet is attached to three metal pads. The device is fabricated in a material stack that consists of four layers: the metal layer, chemical vapor deposition (CVD)-grown monolayer graphene, 300 nm thermally grown silicon dioxide ($SiO_2$), and silicon (Si) wafer.

Highly-stressed metals allow us to induce strain in a controllable and reproducible way. The stress can be induced using two methods. One method is to use forming gas annealing. In this method, 10 nm chromium (Cr) and 140 nm gold (Au) were first deposited by electron beam evaporation and then subjected to forming gas annealing (300 °C, 30 min). The resultant strain is ~530 MPa, measured by laser scanning. By scanning the surface of a wafer before and after metal deposition, we can derive the curvature change of the wafer by calculating the displacement of the reflected beam [19].

Another way we applied to induce strain is to use the internal stress of Cr layer. In this method, four layers of metal consisting of 10 nm Cr, 70 nm Au, 40 nm Cr and 70 nm Au are deposited sequentially, and the resultant internal stress is measured to be ~500 MPa. The amount of strain can be easily controlled by modifying the thickness of Cr film between two Au layers [20,21]. It would be desirable for the entire fabrication flow that eliminates the step of thermal annealing and the requirement for a vacuum furnace. This strain engineering method of harnessing internal stress in a film has been studied in other materials [13,22–25], but not in graphene.

The CVD graphene was patterned using electron beam lithography into a triaxial structure, with three narrow neck regions (Fig. 1(a) red dashed box). The grain size of the CVD-grown monolayer graphene is around 80 μm, and we confirmed that our devices with the largest size are within one grain boundary, thereby excluding any possibility of grain boundary-induced strain effects. Once the underlying $SiO_2$ is etched away, the released stressed metal film shrinks in size, stretching the graphene sheet into three directions, creating the out-of-plane pseudo-magnetic fields. Further fabrication details are described in the supplementary material. Fig. 1(c) presents a schematic illustration of a typical triaxially strained graphene device configuration. The patterned graphene sheet is mechanically pulled along three crystallographic directions by stressed metal pads, and the induced non-uniform strain creates pseudo-magnetic fields [9,17,26] that are illustrated by the out-of-plane arrows. The time-reversal symmetry is preserved without the application of an external magnetic field, which gives rise to pseudo-magnetic fields of opposite signs in the K and K′ valleys [10–12,27–29].

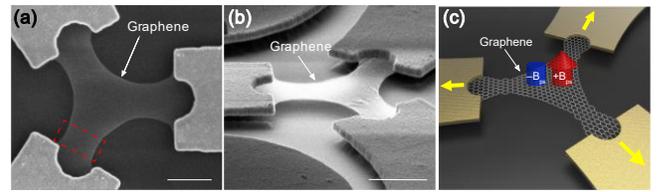

**Fig. 1.** (a, b) Scanning electron micrographs showing fully suspended graphene samples contacted with metal pads. The graphene membranes are suspended about 200 nm above the $SiO_2$ layer. (a) Top-view. Scale bar, 2 μm. (b) Tilted-view. Scale bar, 2 μm. (c) Schematic of a typical triaxially strained graphene device deformed by three metal stressor pads, attaining a uniform pseudo-magnetic field in the center.

We conducted Raman scans to infer the strain using a 532 nm laser source. Figure 2(a) presents the Raman spectra measured for graphene at the center of the suspended membrane and at the most strained neck region, showing a clear strain-induced shift. The strain value can be derived by using a 2D peak strain-shift coefficient of −65.4 $cm^{-1}$/% [30]. The measured strain values for Raman spectra at the neck and the center regions of a triaxially strained graphene are 0.41% and 0.13%, respectively. It is found that the strain at the neck region decreases upon increasing the neck width, and there is a trade-off between the active area of the device and the amount of strain.

Figure 2(b) shows a simulated strain distribution performed by finite-element method (FEM) mechanical simulations. The experimental parameters of the device shown in Fig. 1(a) are used for the structural dimensions and residual stress in the metal pads. The simulated strain values for the center and the neck are 0.18% and 0.39%, respectively. Figure 2(c) compares the calculated (solid line) and experimentally measured (dots) one-dimensional (1D) strain distributions along three crystallographic directions. The three directions are highlighted as dotted lines in inset. The calculated curve is in reasonable agreement with our experimentally measured strain values. The discrepancy between simulation and experimental results may be attributed to the unintentional doping in the CVD graphene layer [31], which may lead to a microscopic spatial variation in the measured Raman shift values. The strain distributions show a gradually increasing strain trend towards the neck region of the triaxial structure. As theoretically proposed in Ref. 9, this structure with a gradually changing strain distribution can induce uniform pseudo-magnetic fields that lead to a large variety of new physical phenomena including zero-field quantum Hall effect [9] and pseudo-Landau level lasers [26].

Figure 2(d) presents a simulated spatial distribution of pseudo-magnetic fields in our experimentally fabricated structure. The gauge potential, $\vec{A}$, is created by the in-plane strain in our structure and can be related to the strain tensors as in the following expression [9,26]:

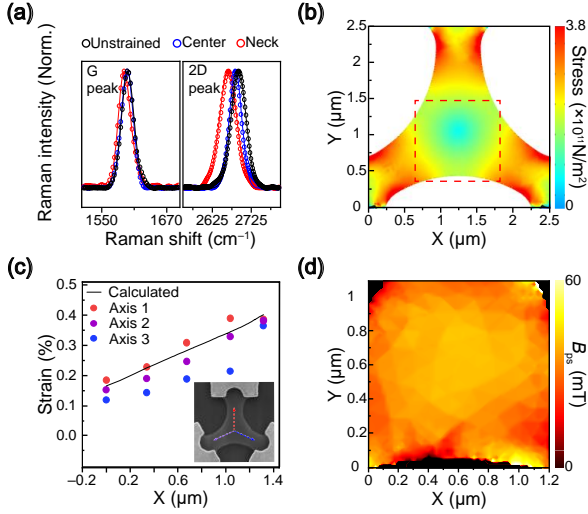

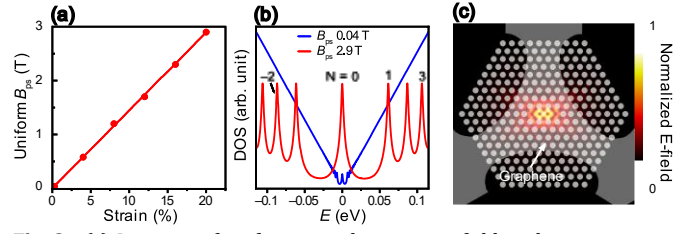

**Fig. 2.** (a) Raman spectra at the neck region and the center of a triaxially strained graphene device. Symbols are measurement data; curves are fitting data. (b) Simulated stress distribution in a graphene device calculated by the FEM simulation. (c) Experimentally determined line cut of 2D Raman peaks along three dashed arrows shown in the inset. The theoretically calculated result (solid line) is also presented for comparison. (d) Calculated pseudo-magnetic field for the boxed region in (b).

$$\vec{A} = (A_x, A_y), \quad A_x = \frac{\beta}{2a_0}(\varepsilon_{xx} - \varepsilon_{yy}), \quad A_y = \frac{-\beta}{a_0}(\varepsilon_{xy}), \quad (1)$$

where $\beta$ is a constant connecting hopping energy and bond lengths [32], $a_0$ is the bond length constant [26,32], $\varepsilon_{xx}, \varepsilon_{yy},$ and $\varepsilon_{xy}$ are the in-plane strain tensor elements [26,32]. Pseudo-magnetic fields, $B_{ps}$, can also be calculated by using the following relation between the gauge potential and pseudo-magnetic fields [9]:

$$B_{ps} = \partial_x A_y - \partial_y A_x. \quad (2)$$

More details on the calculation of pseudo-magnetic fields can be found in Refs 9 and 26. As shown in Fig. 2(d), the spatial distribution of pseudo-magnetic fields can be uniform over a large scale with a maximum field strength of ~0.04 T.

The strength of pseudo-magnetic fields is mainly determined by the intensity of the strain gradient, which is clearly evidenced by Eqs. (1) and (2). The strain gradient can be further increased by inducing a higher maximum strain in the neck region while keeping the size of the strained graphene structure. The maximum strain in the neck region can be conveniently tuned by changing the undercut length of the metal stressing pads, $L_{metal}$, which is defined in the supplementary Fig. S1.

To show the possibility of achieving stronger pseudo-magnetic fields in our proposed structure, we performed further FEM simulations on the same triaxially strained structure for various values of the maximum strain in the neck region. Figure 3(a) shows the strength of the uniform pseudo-magnetic field at the center of graphene as a function of strain in the neck region. The strength of pseudo-magnetic fields reaches up to 2.9 T when the strain in the neck region is assumed to be 20% that is an experimentally achieved level [33]. This unique ability of our structure to tune the strength of pseudo-magnetic fields allows creating distinct graphene optoelectronic devices with different Landau-quantized energy gaps on a single die. Figure 3(b) displays the density of states curve) and 0.04 T (blue curve). The pseudo-Landau level peaks arise under the influence of pseudo-magnetic fields, which could modify the optical response of graphene optoelectronic devices. The detailed calculation procedure of the DOS is provided in Ref. 26. Our triaxially strained graphene can also be used to realize graphene Landau-level lasers [26,34]. Figure 3(c) displays a 3D FDTD simulation result showing a localized optical field at the center of a 2D photonic crystal cavity (PCC). The base structure is composed of a 3-μm-thick silicon slab with a lattice of holes. The hole radius and the periodicity are 2.1 μm and 6 μm, respectively. Triaxially strained graphene is also schematically drawn on top of the cavity. The central area of strained graphene possesses highly uniform pseudo-magnetic fields as shown in Fig. 2(d) and is spatially overlapped with the strong optical field. For a wavelength of 20.6 μm, a quality factor of ~10,000 can be realized according to the simulation. As proposed in Ref. 26, this highly practical structure can be used to achieve population inversion and significant optical net gain in graphene, thus allowing the realization of strained graphene Landau-level lasers.

**Fig. 3.** (a) Intensity of uniform pseudo-magnetic field at the center as a function of the maximum strain in the graphene neck region. (b) The calculated DOS of graphene with a pseudo-magnetic field intensity of 0.04 T and the largest pseudo-magnetic field of 2.9 T shown in (a). (c) FDTD simulation of PCC for the largest pseudo-magnetic field shown in (a).

The current-voltage (I–V) relation for suspended monolayer graphene at zero back-gate voltage has been plotted in Fig. 4. We apply a bias voltage between the two of three electrodes and form a current channel, giving rise to $I_1$ and $I_2$ that correspond to the currents along the blue and red arrows as illustrated in the inset. The similar I–V behavior of the two channels imply that all electrodes possess the same capacity to function as both stressors and metal contacts for current injection. It is worth noting that this triaxial platform could offer a new way to detect the valley Hall effect. Since the K- and K′-valley fermions are separated by pseudo-magnetic fields, carriers located in opposite valleys could accumulate on different edges of the triaxial graphene membrane and move along two channels separately [12]. An imbalance between K and K′ valley can be experimentally imaged under the excitation of a circularly polarized light over a micrometer-scale central active region [35].

In conclusion, we have presented an experimental demonstration of triaxially strained suspended graphene structures. Our unique design may allow obtaining large-area, uniform pseudo-magnetic fields, which can play major roles in harnessing pseudo-magnetic fields for a new class of graphene-based optoelectronic devices. Our device can be utilized to realize practical valleytronic devices including valleytronic transistors [36] and valley filters [12,37]. In addition, the ability to tune the strength of pseudo-magnetic fields via conventional lithography should enable the creation of distinct graphene optoelectronic devices with different Landau-quantized energy gaps on a single die. The presented structure is expected to be very general and can be employed to induce customized strain in any 2D material. We believe that our device lays the groundwork for bringing the performance of graphene-based optoelectronics to another level.

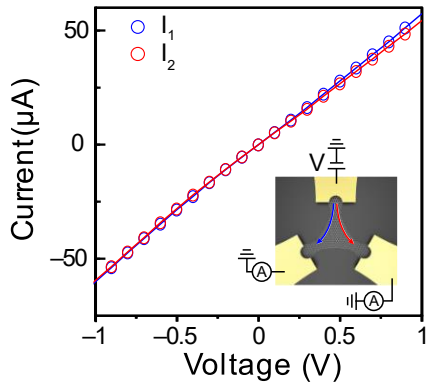

**Fig. 4.** Corresponding I–V curves measured at a zero back-gate voltage. The inset shows a schematic of our device with two experimentally measured current channels highlighted in blue and red arrows.

**Funding.** The research of the project was in part supported by Ministry of Education, Singapore, under grant AcRF TIER 1 2019-T1-002-050 (RG 148/19 (S)). The research of the project was also supported by Ministry of Education, Singapore, under grant AcRF TIER 2 (MOE2018-T2-2-011 (S)). This work is also supported by National Research Foundation of Singapore through the Competitive Research Program (NRF-CRP19-2017-01). This work is also supported by National Research Foundation of Singapore through the NRF-ANR Joint Grant (NRF2018-NRF-ANR009 TIGER). This work is also supported by the iGrant of Singapore A*STAR AME IRG (A2083c0053).

**Acknowledgments.** The authors gratefully thank M. M. Deshmukh for the insightful discussions.

**Disclosures**. The authors declare no competing interests.

**Supplemental document.** See Supplement 1 for supporting content.

# Supplementary Material

for "Triaxially strained suspended graphene for large-area pseudo-magnetic fields"

**Manlin Luo**[1,+] **, Hao Sun**[1,2,+] **, Zhipeng Qi**[3]**, Kunze Lu**[1]**, Melvina Chen**[1]**, Dongho Kang**[1,4]**, Youngmin Kim**[1]**, Daniel Burt**[1]**, Xuechao Yu**[1]**, Chongwu Wang**[1]**, Young Duck Kim**[5]**, Hong Wang**[1]**, Qi-Jie Wang**[1,2]**, and Donguk Nam**[1, *]

[1]School of Electrical and Electronic Engineering, Nanyang Technological University, 50 Nanyang Avenue, Singapore 639798, Singapore
[2]Division of Physics and Applied Physics, School of Physical and Mathematical Sciences, Nanyang Technological University, 21 Nanyang Link, Singapore 637371, Singapore
[3]School of Physics and Optoelectronic Engineering, Nanjing University of Information Science & Technology, Nanjing 210044, China
[4]School of Electrical Engineering and Computer Science, Gwangju Institute of Science and Technology, Gwangju 61005, Republic of Korea
[5]Department of Physics, Kyung Hee University, Seoul 02447, Republic of Korea
* email: dnam@ntu.edu.sg
+these authors contributed equally to this work

## Contents

**S1. Fabrication procedure**

To make triaxially strained suspended graphene, we use chemical vapor deposition (CVD) monolayer graphene membranes prepared on an oxidized silicon wafer (285 nm of $SiO_2$), provided by Grolltex. Before defining triaxial graphene arrays using the VISTEC EBPG5200 electron-beam lithography (EBL) system, polymethyl methacrylate (PMMA, 950K A6) is spin coated on the graphene membrane at a speed of 4500 rpm for 90 s, followed by soft baking at 180°C for 2 minutes. The PMMA on an area adequate to fit in the electrodes around the triaxial graphene structure is exposed and developed. Unexposed PMMA on graphene acts as etch masks in the subsequent reactive ion etching process. Electrodes are patterned using the same PMMA spin coating and EBL exposure recipes. Then the metal layer comprising Cr and Au is deposited by electron beam evaporation. The residual stress in metal is induced by forming gas annealing (300 °C, 30 min) or internal stress of Cr, which is measured to be 500~530 MPa. In order to make the triaxial graphene membrane suspended, the sample is rinsed in 5% HF, removing the $SiO_2$ under graphene, after which the sample is transferred into deionized water and isopropanol. Followed by critical point drying.

**S2. Strain engineering of suspended graphene**

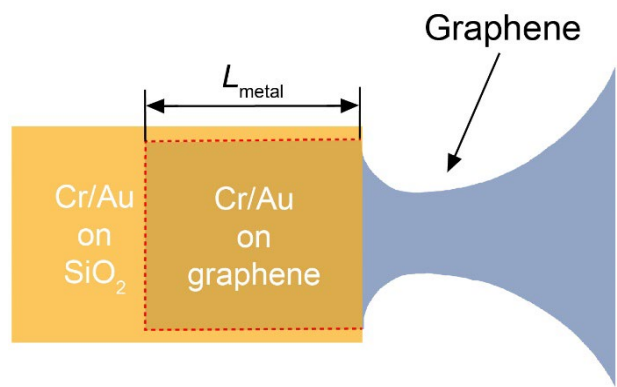

Fig. S1. Schematic illustration of a metallic contact on graphene.

After graphene nano-patterning, the metallic contact is defined by electron-beam lithography and deposited by electron beam evaporation. The graphene is then suspended by HF etching of the $SiO_2$ layer and drying. The metal layer in this study serves as stressors, electrodes and clamps. We can release the Cr/Au on graphene rapidly during the etching process because of the poor adhesion between graphene and $SiO_2$. In the meantime, HF solution penetrates the $SiO_2$ under Cr/Au with a low transversal etching rate (<10 nm/min). Therefore, the Cr/Au on $SiO_2$ functions as clamps while Cr/Au on graphene functions as stressors. The maximum strain in the neck region can be easily controlled by customizing the undercut length of the metal stressing pads, $L_{\text{metal}}$ as illustrated in the schematic. And the strain gradient can be increased by inducing a higher maximum strain in the neck region while keeping the active region of strained graphene unchanged.